# Polar Discontinuity Doping of the $LaVO_3/SrTiO_3$ Interface

Y. Hotta,[1,a] T. Susaki,[1] and H. Y. Hwang[1,2,b]

[1]Department of Advanced Materials Science, University of Tokyo, Kashiwa, Chiba 277-8651, Japan

[2]Japan Science and Technology Agency, Kawaguchi, 332-0012, Japan

## Abstract

We have investigated the transport properties of $LaVO_3/SrTiO_3$ Mott insulator/band insulator heterointerfaces for various configurations. The (001)-oriented n-type $VO_2/\mathit{LaO}/TiO_2$ polar discontinuity is conducting, exhibiting a $LaVO_3$ thickness-dependent metal-insulator transition and low temperature anomalous Hall effect. The (001) p-type $VO_2/\mathit{SrO}/TiO_2$ interface, formed by inserting a single layer of bulk metallic $SrVO_3$ or SrO, drives the interface insulating. The (110) heterointerface is also insulating, indicating interface conduction arising from electronic reconstructions.

The finding of a metallic, conducting layer at the interface between the wide band gap perovskite insulators $LaAlO_3$ and $SrTiO_3$ [1] has stimulated much recent activity. At the atomic scale, this interface between the alternately charged atomic layers in $LaAlO_3$ and the charge-neutral atomic layers in $SrTiO_3$ produces an electrostatic potential which would diverge with $LaAlO_3$ thickness, in the absence of any reconstructions [2]. The emerging picture is that this "polar catastrophe" can be resolved using very different mechanisms than observed in conventional semiconductor heterointerfaces [3,4]. Unlike semiconductors, transition metal cations near the interface can adopt a mixed-valence ionic state, whose charge balances the polar discontinuity [2,5,6]. These studies indicate that polar discontinuities can be used to create unusual interface electronic states using electrostatic boundary conditions [7-9].

Although the discussion above is general, most recent experimental studies have focused on the $LaAlO_3/SrTiO_3$ interface [10,11], although theoretical investigations have been performed for other candidate interfaces [12,13]. In order to examine whether this phenomenon is general, or unique to the $LaAlO_3/SrTiO_3$ interface, we have investigated the $LaVO_3/SrTiO_3$ interface. At the ionic level, this interface presents the same polar discontinuity, in that $LaVO_3$ is composed of charged stacks of $(LaO)^{+}$ and $(VO_2)^{-}$ in the (001) direction, while $SrTiO_3$ is composed of charge neutral stacks. There are of course notable differences: $LaVO_3$ is a Mott insulator [14] – that is, a system which conventional band theory would predict to be a metal, but where strong electron-electron interactions create a correlated insulating state. Furthermore, vanadium oxides can be formed with a range of formal valence states from 2+ to 5+. By contrast, $LaAlO_3$ has relatively few charge degrees of freedom.

Making use of our previous studies of the $LaVO_x$ thin film growth phase diagram [15], we have investigated the transport properties of three epitaxial $LaVO_3/SrTiO_3$ interfaces: 1) The (001) oriented $VO_2$/*LaO*/$TiO_2$ interface, which has an n-type polar discontinuity which requires a net charge of $-e/2$ (where $e$ is the fundamental charge) to resolve the polar catastrophe; 2) The (001) oriented $VO_2$/*SrO*/$TiO_2$ interface, which has a p-type discontinuity requiring $+e/2$ net charge; and 3) The (110) heterointerface, which has no ionic polar discontinuity and thus requires no significant charge reconstruction [16]. Of the three, only the first interface is found to be conducting and shows metallic behavior. Furthermore, a minimum thickness of 5 unit cells (uc) of $LaVO_3$ is required to form a conducting n-type interface. The thickness dependent transport properties and low temperature anomalous Hall effect of the n-type interface suggests coupling of the interface electrons to the Mott insulator $LaVO_3$.

$LaVO_3$ thin films were grown on $SrTiO_3$ substrates by pulsed laser deposition (PLD) using a $LaVO_4$ polycrystalline target. Most of the structures were grown at 600 $^oC$ under an oxygen partial pressure of 1 x $10^{-6}$ Torr, with a laser fluence of 2.5 J/cm$^2$, following our previous optimization for two-dimensional layer-by-layer growth of $LaVO_3$ [15]. By depositing $LaVO_3$ directly on $TiO_2$ terminated (001)-oriented $SrTiO_3$ substrates, the n-type interface 1) is formed, as shown in Fig. 1 (a). The p-type interface 2) can be formed by growing a single layer of SrO on $TiO_2$ terminated (001)-oriented $SrTiO_3$ substrates using a SrO single crystal target [1]. For the interface series using SrO to vary the interface, a higher laser fluence of ~3 J/cm$^2$ was used, to accommodate the SrO formation. Alternatively, we can take advantage of the multiple valence character of vanadium, and insert a single perovskite unit cell of $SrVO_3$ using a polycrystalline $Sr_2V_2O_7$ target before depositing $LaVO_3$, as shown in Fig. 1 (c). Although similar results were obtained with both approaches, the second approach takes advantage of the enhanced stability of perovskite unit cell deposition by PLD. Finally, (110)-oriented substrates were used to grow a control interface 3) free of a polar discontinuity. The films were grown in the layer-by-layer growth mode and monitored by reflection high-energy electron diffraction (RHEED) oscillations, as shown in Fig. 1 (b) and (d). The resulting films, probed by atomic force microscopy, were atomically smooth and exhibited unit cell steps reflecting the slight miscut angle of the substrate.

Figure 2 (a) summarizes the thickness dependence of the temperature dependent resistivity $\rho(T)$ for the n-type interface 1), plotted here as a three-dimensional resistivity $\rho_{3D}$ as normalized by the film thickness $t$. Ohmic contacts were formed by wire bonding or ultrasonic soldering in 6 probe Hall bar geometry, which penetrated through the interface. All films for $t \leq 4$ uc in thickness were insulating ($t < 4$ could not be measured), whereas thicker films exhibited metallic behavior. In contrast to the wide range of values found for $\rho_{3D}$, the same data plotted as $\rho_{2D}$ (Fig. 2 (b)) shows that the data for all of the metallic samples essentially collapse to a narrow range of $\rho_{2D}$ values. This indicates that the $LaVO_3$ film itself is indeed insulating, and that the interface forms the conducting channel.

For $LaAlO_3/SrTiO_3$ interfaces, several groups have reported that the conducting interfaces they have studied may arise from growth induced oxygen vacancies in the $SrTiO_3$ substrate [17,18]. This is particularly important to address for the $LaVO_3/SrTiO_3$ interfaces studied here, since oxygen post-annealing is generally unavailable due to the further oxidation of the film (forming $LaVO_4$). Already the scaling of $\rho_{2D}$ shown in Fig. 2 (b) rules out a doping mechanism that scales with exposure to the kinetics of film growth.

Further direct evidence for an intrinsic doping mechanism via the polar discontinuity can be found in Fig. 2 (c), where $\rho_{2D}$ for the p-type (001) interface 2) and the (110) interface 3) exhibit insulating behavior. That the p-type interface is insulating is quite striking: this was grown by inserting 1 uc of $SrVO_3$, a good metal in bulk. Given that insertion of this metal layer converted a metallic interface to an insulating one gives strong evidence for analyzing the interfaces in terms of reconstructions in response to polar discontinuities, not growth induced oxygen vacancies. Although the (110) orientation is a polar direction, any two perovskites can be joined across a (110) interface without introducing an ionic polar discontinuity. This control interface exhibits the most insulating behavior of all the configurations measured. These results also rule out conductivity arising from significant interdiffusion, forming $(La,Sr)TiO_3$ and $(La,Sr)VO_3$ near the interface (both conductors in bulk). Therefore, we can conclude that the transport properties of the interface critically depend on the polar structure of the interface.

Figure 3 presents an alternative approach to vary between the n-type interface 1) and the p-type interface 2), by gradually inserting a monolayer of SrO. Here again a metal-insulator transition is observed, indicating the intrinsic insulating nature of the p-type interface 2), independent of the growth approach or surface termination. Note that the two experimental approaches to forming p-type interfaces give alternative $LaVO_3$ surface terminations – $VO_2$ termination for the case of $SrVO_3$ insertion, and LaO termination for the case of SrO insertion. This insulating state is consistent with an electronic reconstruction using $V^{4+}$ states, but with a peak density below the mobility threshold: bulk $(La,Sr)VO_3$ requires ~20% hole doping for metallic behavior [14]. Alternatively, oxygen vacancies may still provide an energetically favorable resolution for the polar discontinuity, as found at the p-type $LaAlO_3/SrTiO_3$ interface [2,6]. Spectroscopic investigation of the interface will be necessary to resolve between the two scenarios.

We next examine the thickness dependence of the conducting n-type interface 1), given in Fig. 4. The insulator to metal transition occurs between 4 and 5 uc thick $LaVO_3$, in close proximity to crossing $h/e^2 \sim 25.8$ kΩ (where $h$ is Plank's constant), the resistance quantum threshold for two-dimensional transport [19]. Approaching this transition, the carrier density $n$ rapidly decreases from a large thickness limiting value of $10^{14}$ $cm^{-2}$, decreasing by over two orders of magnitude (100 K value) between $t = 10$ uc and $t = 4$ uc. Remarkably, the Hall mobility $\mu$ is significantly enhanced in this regime (by a factor of 5 at 100 K), indicating an insulator-metal transition driven by density, not mobility. The length scale for the transition is similar to that reported previously for $LaAlO_3/SrTiO_3$, for the case of coupling n-type and p-type interfaces [3], or for coupling to a polar surface [4]. This length scale can be interpreted as the threshold above which reconstructions are energetically favored, and below which the dipole shift induced by the finite thickness of the polar material is sustainable [20].

The analysis of Fig. 4 was taken at relatively high temperatures due to our finding of an unusual low temperature anomalous Hall effect. Figure 5 (a) shows the temperature dependent Hall resistance up to 14 tesla for the $t = 1000$ uc n-type interface 1). At and above 100 K, the Hall response is linear. At lower temperatures, however increasing curvature can be observed at high fields (~ 9 tesla). The anomalous Hall effect is conventionally associated with magnetism or scattering from magnetic impurities [21], although recently more exotic mechanisms arising from topological features in the real space spin structure [22], or reciprocal space electronic structure [23] have been much discussed.

Although the origin of anomalous Hall effect at the $LaVO_3/SrTiO_3$ interface is presently unclear, we suggest that the data of Figs. 4 and 5 can be interpreted by considering the evolution of the charge distribution at the interface with *t*. In contrast to $LaAlO_3/SrTiO_3$ n-type interfaces, the barrier for electrons in the $LaVO_3$ can be expected to be significantly lower (i.e., possible admixture of $V^{2+}$ states). Thus the charge density may have nontrivial weight in $LaVO_3$ as well as in $SrTiO_3$, and couple to the antiferromagnetic, orbitally-ordered ground state of $LaVO_3$ [14]. In particular, the conducting electrons at the interface are particularly sensitive to the spin configurations at the surface termination of the bulk antiferromagnet. In pyrochlore molybdates, the anomalous Hall effect is induced by spin textures formed on the geometrically frustrated spin lattice [22,24]; here, the Hall resistance probes the surface/interface spin structure. As the thickness of $LaVO_3$ is reduced, the charge density on the $LaVO_3$ side is also reduced – the mobility enhancement for $t < 10$ seen in Fig. 4 (c) is consistent with the higher electron mobility of $SrTiO_3$ relative to $LaVO_3$. Also consistent is our finding that the anomalous (nonlinear) Hall effect is suppressed with decreasing thickness, such that by $t = 5$, a linear Hall response is recovered (Fig. 5 (b), raw data shown in the inset).

In summary, these studies of configuration-dependent conductivity at the $LaVO_3/SrTiO_3$ interface demonstrate that polar discontinuity doping by electronic reconstructions can be observed at the interface with a Mott insulator, as well as between band insulators as in $LaAlO_3/SrTiO_3$. They further suggest that interface engineering is an attractive, experimentally tractable approach to creating novel two-dimensional states in correlated electron systems, as proposed theoretically for a number of artificial structures [13,25].

We thank A. Fujimori, H. Takagi, and N. Nagaosa for helpful discussions. We acknowledge support from a Grant-in-Aid for Scientific Research on Priority Areas. Y.H. acknowledges partial support from QPEC, Graduate School of Engineering, University of Tokyo.

[a]Present address: Institute of Scientific and Industrial Research, Osaka University, 8-1 Mihogaoka, Ibaraki, Osaka 567-0047, Japan

[b]To whom correspondence should be addressed. Electronic address: hyhwang@k.u-tokyo.ac.jp

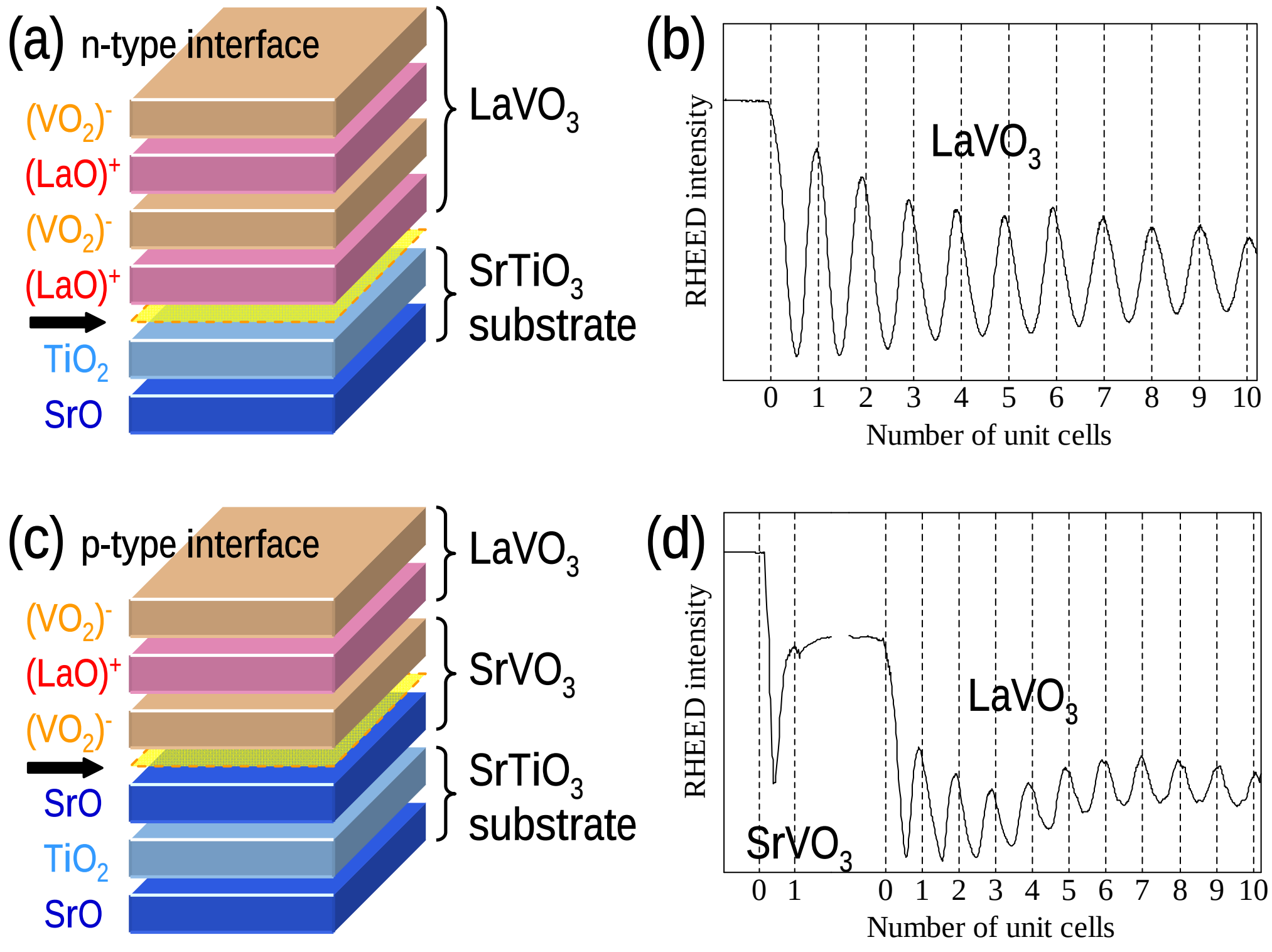

FIG. 1. Schematic illustrations and growth of two possible configurations of the interface between $LaVO_3$ and $SrTiO_3$ in the (001) orientation. (a) Schematic of the (001) $VO_2$/*LaO*/$TiO_2$ interface, inducing an n-type polar discontinuity. (b) RHEED intensity oscillations for the growth of $LaVO_3$ on the $TiO_2$-terminated (001) $SrTiO_3$ substrate. (c) Schematic of the (001) $VO_2$/*SrO*/$TiO_2$ interface, inducing a p-type polar discontinuity. (d) RHEED oscillations for the growth of 1 uc $SrVO_3$ followed (with a ~25 second delay) by $LaVO_3$ on the $TiO_2$-terminated (001) $SrTiO_3$ substrate. Arrows in (a) and (c) denote the chemical interface.

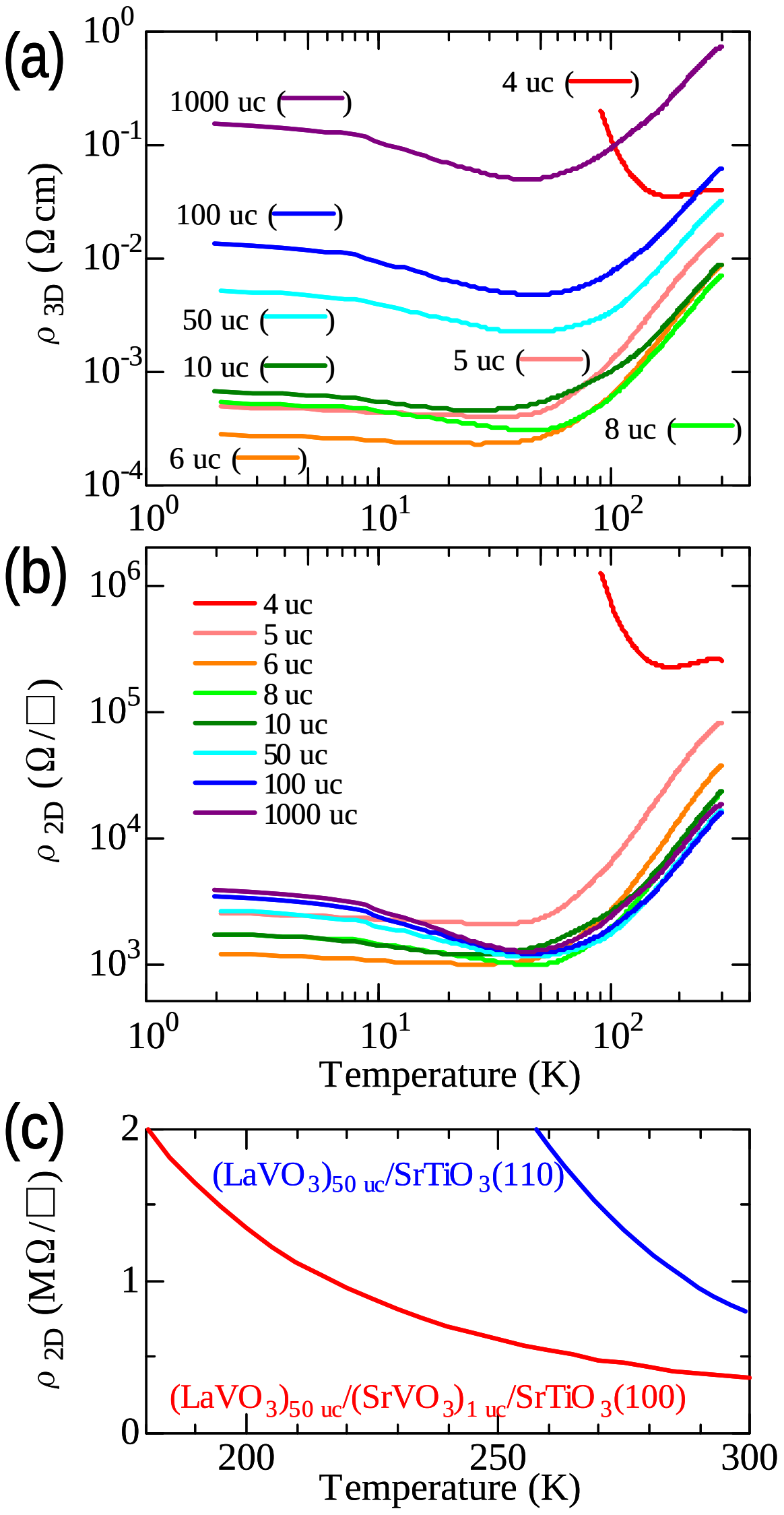

FIG. 2. Temperature dependent resistivity $\rho(T)$ of the $LaVO_3/SrTiO_3$ structure for various configurations. Three-dimensional resistivity $\rho_{3D}$ (a) and two-dimensional resistivity $\rho_{2D}$ (b) for a variety of $LaVO_3$ thicknesses with the (001) $VO_2$/*LaO*/$TiO_2$ n-type interface. (c) $\rho_{2D}(T)$ for configurations of the (001) $VO_2$/*SrO*/$TiO_2$ p-type interface [using $(LaVO_3)_{50\ uc}/(SrVO_3)_{1\ uc}/SrTiO_3$ (001)] and the non-polar (110) $(LaVO_3)_{50\ uc}/SrTiO_3$ interface.

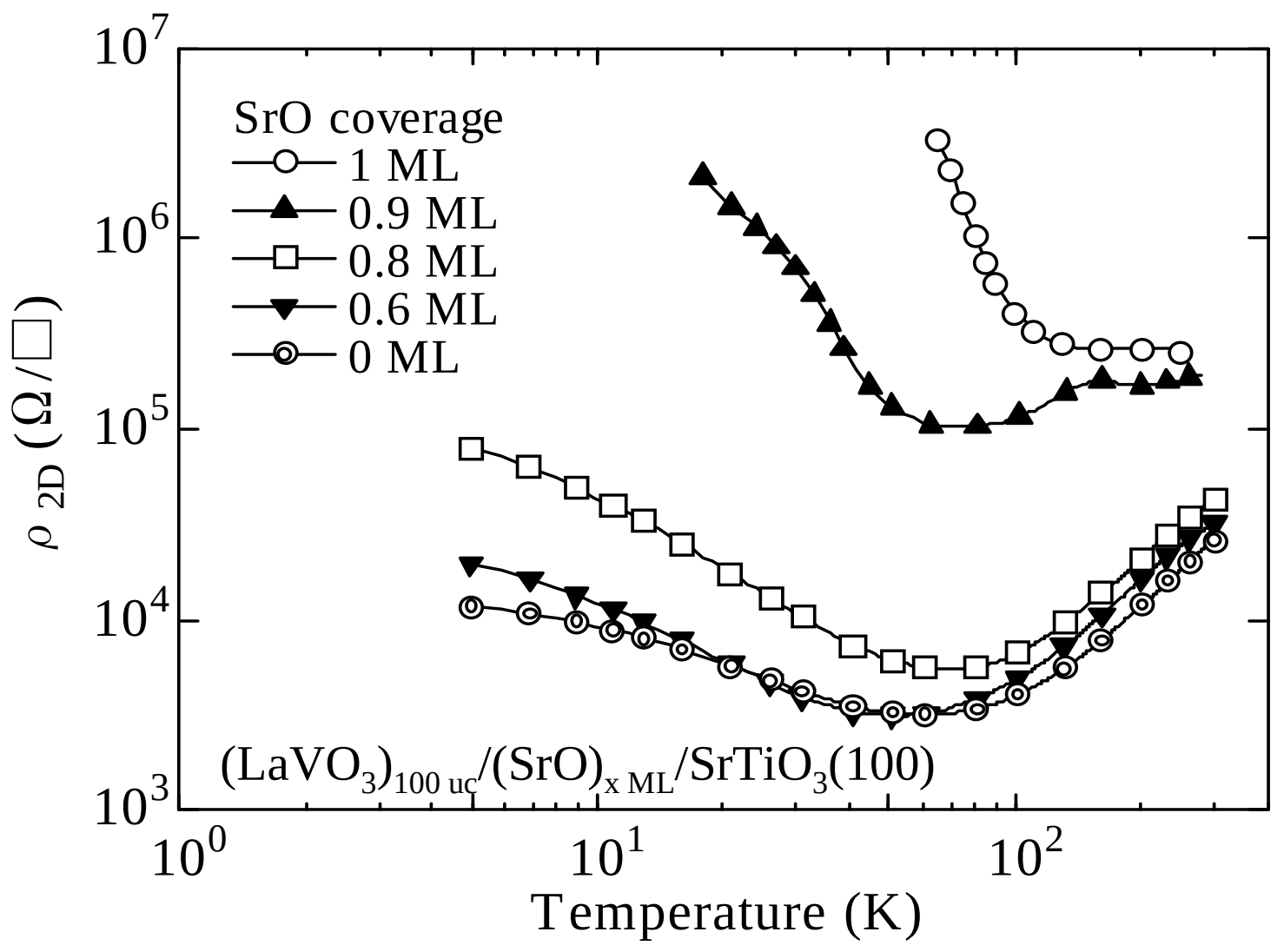

FIG. 3. $\rho_{2D}(T)$ for $(LaVO_3)_{100\,uc}/SrTiO_3$ (001), for variable insertion of 0 to 1 monolayer of SrO, going from the n-type to the p-type interface.

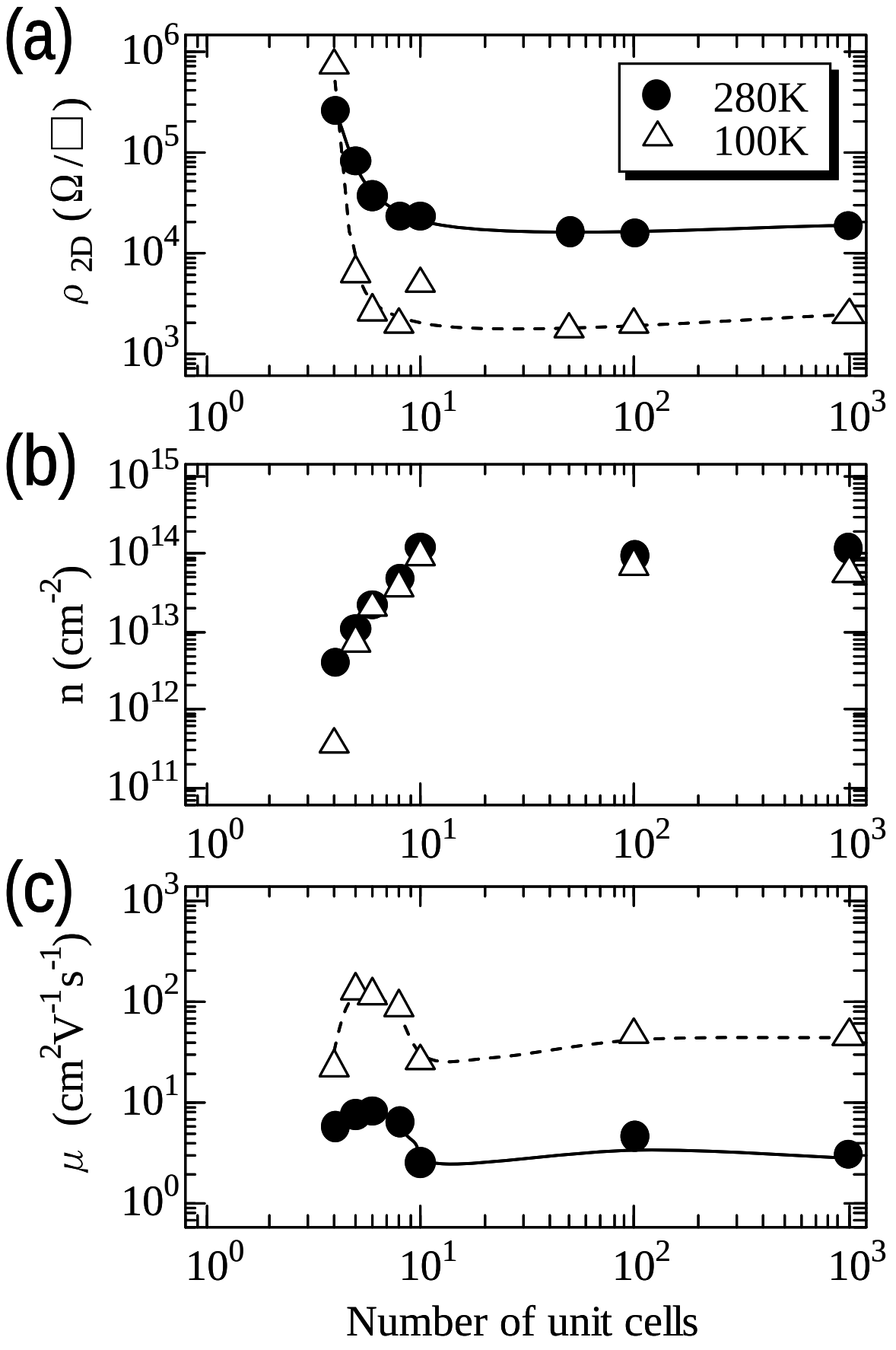

FIG. 4. $LaVO_3$ thickness dependence of the transport properties for the (001) $VO_2$/*LaO*/$TiO_2$ n-type interface at the measurement temperatures of 280 K and 100 K. The thickness dependence of $\rho_{2D}$ (a), carrier density $n$ (b), and mobility $\mu$ (c) are given. Solid and dashed lines are guides to the eye.

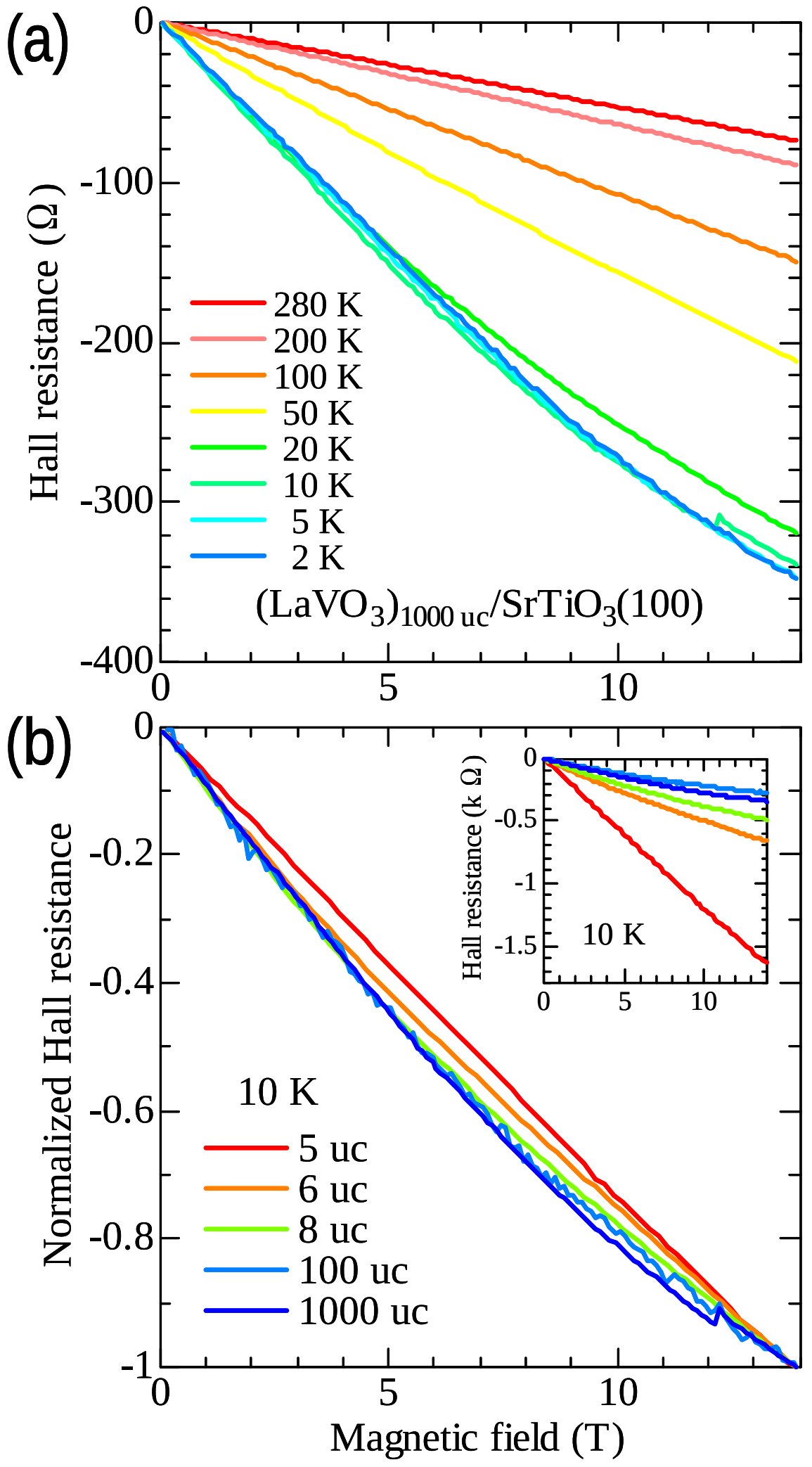

FIG. 5. Magnetic field and temperature dependence of Hall resistance $R_{xy}(T,H)$. (a) $R_{xy}(T,H)$ of the (001) $VO_2$/*LaO*/$TiO_2$ n-type interface for $(LaVO_3)_{1000}$ uc/$SrTiO_3$(100). (b) $R_{xy}(H)$ normalized to the 14 tesla value for different thicknesses of $LaVO_3$ at the measurement temperature of 10 K. The inset shows the original $R_{xy}(H)$ data before normalization.